# Physics-based compact modeling for the drain current variability in single-layer graphene FETs

Nikolaos Mavredakis, Anibal Pacheco-Sanchez, Ramon Garcia Cortadella, Anton-Guimerà-Brunet,
Jose A. Garrido, and David Jiménez

*Abstract*— **For the growth of emerging graphene field-effect transistor (GFET) technologies, a thorough characterization of on-wafer variability is required. Here, we report for the first time a physics-based compact model which precisely describes the drain current ($I_D$) fluctuations of monolayer GFETs. Physical mechanisms known to generate *1/f* noise in transistors such as carrier number and Coulomb scattering mobility fluctuations are revealed also to cause $I_D$ variance. Such effects are considered in the model by being activated locally in the channel and the integration of their contributions from source to drain results in total variance. The proposed model is experimentally validated from a statistical population of three different-sized solution-gated GFETs from strong p- to strong n-type bias conditions. A series resistance $I_D$ variance model is also derived mainly contributing at high carrier densities.**

## I. INTRODUCTION

VARIABILITY in transistor technology can be classified into global and local variation components. The former includes statistical fluctuations outside the actual chip/die (lot-to-lot, wafer-to-wafer, die-to-die) while the latter contains within-die device-to-device variations. The sum of all these components defines the total variance [1, §2]. State-of-the-art subnanometer sized conventional silicon (Si) technology is constrained due to scaling effects, which has led to the study of emerging two-dimensional devices such as graphene field-effect transistors (GFETs). Outstanding intrinsic properties of graphene such as immense carrier mobility and saturation velocity have established GFETs as suitable for high-speed analog/RF applications [2], [3]; graphene evidently presents superior intrinsic characteristics to other state-of-the-art materials (Si, GaN, GaAs) widely employed in high-frequency FETs (cf. [2, Table 1]). Additionally, cut-off frequencies near 500 GHz have been measured for GFETs, close enough to the maximum value (688 GHz) of well-established III-V HEMTs [3]. GFET maximum oscillation frequencies are yet quite below HEMTs but already in the range of 200 GHz which is quite impressive for a material still in its embryonic stage [3]. Additionally, graphene surpasses its competitors in flexible electronics due to its high strain limits [3]. GFETs are also ideal for biomedical sensing applications [4] where the operating principles of such sensors are mainly based on GFET signal fluctuations [4], [5]; Electrolyte-gated GFETs have presented excellent behavior in such sensors [4]. Unique opto-electronic properties of graphene have also led to its employment in THz detection applications [6]. For GFET integrated circuits' (ICs) optimum performance, on-wafer fabrication of GFET chips with excellent performance and yield is essential [7]-[9]. Hence, the characterization and modeling of variability in GFETs are critical for enhancing fabrication procedures and wafer yield targeting towards large-scale IC production [5], [7]-[10]. Experimental GFET variability aspects have been mainly investigated in [5], [7]-[9], while in [10] we have proposed and experimentally validated a physics-based compact model for the GFET *1/f* noise variance.

Drain current ($I_D$) variability models have been proposed for advanced Si [11]-[23] as well as organic FETs [24]-[25], where most of them [11]-[21] calculate the $I_D$ variance by differentiating the final compact $I_D$ equation with respect to the critical parameters that contribute to variability (cf. eq. 1 in [15]). Such parameters are mainly threshold voltage $V_{TH}$, low field mobility $\mu$ [11]-[21], series resistance $R_c$ [13]-[15], [17]-[21], mobility degradation coefficient $\Theta_{int}$ [14]-[15] and physical process parameters [16]. According to previous works [11]-[21], $I_D$ variability is generated by the global variations of physical electrical parameters, which are implicitly induced mainly by the average charged impurities $Q_{imp}$ fluctuations over the whole device channel. Such an

This work has received funding from the European Union's Horizon 2020 research and innovation programme under grant agreements No GrapheneCore3 881603, from Ministerio de Ciencia, Innovación y Universidades under grant agreements RTI2018-097876-B-C21(MCIU/AEI/FEDER, UE), FJC2020-046213-I and PID2021-127840NB-I00 (MCIN/AEI/FEDER, UE).
N. Mavredakis, and D. Jiménez are with the Departament d'Enginyeria Electrònica, Escola d'Enginyeria, Universitat Autònoma de Barcelona, Bellaterra 08193, Spain. (e-mail: Nikolaos.mavredakis@uab.cat).
A. Pacheco is with the Departamento de Electrónica y Tecnología de Computadores, Universidad de Granada, 18011 Granada, Spain.
R. Garcia-Cortadella, and J. A. Garrido are with the Catalan Institute of Nanoscience and Nanotechnology (ICN2), CSIC, Barcelona Institute of Science and Technology, Campus UAB, Bellaterra, Barcelona, Spain.
A. G. Brunet is with the Institut de Microelectrònica de Barcelona (IMB-CNM), CSIC, Esfera UAB, 08193 Bellaterra, Spain and Centro de Investigación Biomédica en Red en Bioingeniería, Biomateriales y Nanomedicina (CIBER-BBN), 28029 Madrid, Spain approach is inaccurate especially in non-linear regime, as it neglects local $Q_{imp}$ deviations ($\delta Q_{imp}$) at individual channel sections, which are much stronger than global ones due to their stochastic nature [22]; $\delta Q_{imp}$ include dopant and trap variations in depletion and oxide interface charges, respectively. Such $\delta Q_{imp}$, mainly inducing $V_{TH}$ variations, are the primary $I_D$ variance generators in Si [1, §2], [11], [12, Table II], [22], [23] and organic FETs [24]-[25]. Despite the lack of depletion charge dopants, $\delta Q_{imp}$ remain the main source of $I_D$ variance in still not optimized GFET technologies, due to graphene sensitivity on variabilities arising from its interaction with the dielectric layer and/or the ambient environment [5, Table 1]. As the device channel width ($W$) and length ($L$) shrink, variations primarily caused by line edge (LER) and line width roughness (LWR) disorders, become sizeable and also contribute to $I_D$ variance [1, §2], [11], [12, Table II], [22], [23]. LER, LWR, are out of scope of the present study where the bias dependence of $I_D$ variance in wide-long GFETs is mainly examined since edge effects are imperceptible for such large dimensions.

A physics-based single-layer (SL) GFET $I_D$ variance compact model is for the first time proposed in this work considering local current fluctuations [23]-[25]. Hence, the device channel is divided into infinitesimal slices where each one of them corresponds to a local $I_D$ variance source; such sources are considered uncorrelated. By integrating from source (S) to drain (D), total $I_D$ variance is derived by considering the $\delta Q_{imp}$ contribution to local fluctuations of i) channel transport charge $Q_{gr}$ [23], [25] and ii) effective mobility $\mu_{ueff}$ through Coulomb scattering [25]. As it will be shown, the former mechanism (i) is identical to the carrier number fluctuation $1/f$ noise model [26, §6] as it is the case between [23] and [27], while the latter (ii) is equivalent to the correlated mobility fluctuations $1/f$ noise model [26, §6], [28] likewise as between [25] and [29]. Note that long-range Coulomb scattering induced by impurities, is a dominant mechanism in GFETs [30]-[34] mainly defining $\mu_{ueff}$ near charge neutrality point (CNP) [30]-[32] where short range- and phonon-scattering are less critical due to the low net charge of graphene near CNP; Coulomb scattering mobility $\mu_c$ is $V_{GS}$-independent in SL GFETs [30]-[34], in contrast to incumbent Si technologies where there are claims of a (weak) bias-dependence [28], [35]. Distinct extrinsic and intrinsic Coulomb scattering effects have been reported in some Si FETs $1/f$ noise models [36] where the extrinsic ones include $R_c$ contribution to $1/f$ noise and consequently to $I_D$ variance. This occurs because in such models, $R_c$ impact on $IV$ part is considered together with channel-induced mechanisms (cf. eqs 1-16 in [36]). On the contrary, in our Verilog-A implementation of the model [37]-[40], also used here, $R_D$, $R_S$ ($R_c=R_D=R_S$) are connected as extrinsic resistors to internal D, S, respectively [38], while a simplified separate $I_D$ variance model accounting for $R_c$ contribution, is adopted from Si technologies [17]-[21]. Notice that the impact of $R_c$ on the different ambipolar transport regions has also been considered in the core $I_D$ model here (cf. eq. 4 in [38]). The total $I_D$ variance model derived in the present study, is based on the chemical potential ($V_c$)-, physics-based SL GFET compact model [37]-[40]. All the model features, including the present work, have been tested in a Spice circuit simulator (Keysight ADS) while different proposed circuit topologies have accurately benchmarked the model [3], [37]. Additionally, the IV part of the model has also been implemented in Qucs open access Spice-like simulator [41].

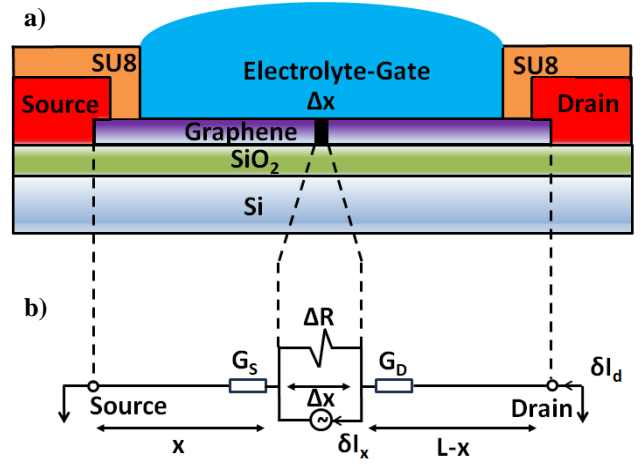

**Fig.1**. a) Fabricated top electrolyte-gated single-layer GFET under test. b) Equivalent small-signal circuit of the split GFET channel.

## II. DEVICES-MODEL DERIVATIONS

To examine the $I_D$ variability in GFETs, arrays of top-, solution-gated (SG) SL GFETs have been employed with $W/L=100/100$ (50 GFETs), $50/50$ (27 GFETs) and $20/20$ (22 GFETs) $\mu m/\mu m$. The device under test (DUT) is presented in Fig. 1a where the graphene channel, the metal contacts, the SU8 passivation layer, the electrolyte top gate and the Si/SiO$_2$ substrate are depicted. Note that a physical top-gate dielectric is not present but a top-gate capacitance of around $\sim 2$ $\mu F/cm^2$ has been measured [10, Table I], corresponding to an equivalent oxide thickness of $1.7$ $nm$ employed for the modeling purposes of the present study. A more detailed discussion on fabrication and measurements conditions can be found elsewhere [4], [10]. $I_D$ has been measured on wafer, from strong p-type to strong n-type bias conditions (top-gate voltage $V_{GS}$ sweep) at drain voltage $V_{DS}= 50$ m$V$.

To derive a physical, $V_c$-based $I_D$ variance model suitable for circuit simulations, the effect of all the local fluctuations along the channel must be accurately formulated. Thus, as in [23], [25], the GFET is sliced into two smaller transistors immune to fluctuations with same $W$ and lengths $x$, $L-x$, respectively, and a microscopic slice $\Delta x$ between them where a current source is employed to model the local current fluctuation $\delta I_x$ connected in parallel with the resistance $\Delta R$ of the channel slice (cf. Fig. 1b); $\delta I_d$ is the current deviation due to local $\delta I_x$. Due to negligible voltage fluctuations across $\Delta R$ [26, §6], [40], small-signal analysis can be applied which permits the replacement of the two transistors with $G_S$, $G_D$ conductances at S and D sides, respectively [23], [25]. The same subcircuit (Fig. 1b), is used for $1/f$ noise modeling methodology in Si FETs [26, Fig. 6.2], [27, Fig. 1], in organic FETs [29, Fig. 1] and in our prior work regarding

GFETs [10, Fig. S1b], consistent with the argument that $1/f$ noise and $I_D$ variance are equivalent effects, as they are generated by identical mechanisms [23]. $I_D$ deviation due to local current fluctuation from $\Delta x$ equals to [25, eq. 3]:

$$\delta I_d = G_{CH}\Delta R \delta I_x = \frac{\Delta x}{L_{eff}}\frac{\mu_{ueff}}{\mu_{diff}}\delta I_x = \frac{\Delta x}{L}\delta I_x \quad (1)$$

with $G_{CH}$ the channel conductance at $\Delta x$ [40]; $\mu_{ueff}$ includes $\mu_c$ [28], degraded mobility due to vertical field ($\mu_u$) [39] and velocity saturation (VS) effects [40], $\mu_{diff}$ is the differential mobility [40] and $L_{eff}$ is the effective channel length due to VS [37, §2.1.2], [40] (see Appendix (a) for eq. 1 derivation).

Next, the relative local current fluctuations due to $\delta Q_{imp}$ must be expressed in terms of $V_c$, and then the total $I_D$ variance is derived by integrating from S to D by using $V_c$ as the integral variable, similarly to the rest of the modules of the GFET compact model [37]-[40]. Hence [25], [28]:

$$\frac{\delta I_x}{I_D} = \frac{\partial Q_{gr(x)}}{Q_{gr(x)}}|\delta Q_{imp} \pm \frac{\partial \beta}{\beta}|\delta Q_{imp} \quad (2)$$

with $\beta = \mu_{ueff}C_{t(b)}W/L$ the transconductance factor and $C_{t(b)}$ the top(back) gate oxide capacitance. The 1st and 2nd term in eq. 2 represent the relative local fluctuations of $Q_{gr}$ and $\beta$ respectively, due to $\delta Q_{imp}$. From IV GFET model definitions [37, §2.1.2]-[40] and charge conservation law [23], [25]:

$$Q_{gr} = \frac{k}{2}\left(V_c^2 + \frac{\alpha}{k}\right) \Leftrightarrow \frac{\partial Q_{gr}}{\partial V_c} = k|V_c| \quad (3)$$

$$Q_{t(b)} = C_{t(b)}(V + V_c - V_G - V_{G0}) \Leftrightarrow \frac{\partial Q_{t(b)}}{\partial V_c} = C_{t(b)} \quad (4)$$

$$\delta Q_{gr} + \delta Q_t + \delta Q_b = \delta Q_{imp} \Leftrightarrow \frac{\partial Q_{imp}}{\partial V_c} = k|V_c| + C \quad (5)$$

where $k(=2e^3/(\pi(hu_f)^2))$ is a constant model coefficient (cf. eq. 4 in [37]) with $e$ the electron charge, $h$ the reduced Planck constant and $u_f$ the Fermi velocity [37]; $\alpha=2n_0e$ a residual charge ($n_0$) related term [40], $Q_{t(b)}$ is the top(back) gate oxide charge and $C=C_t+C_b$ [37, §2.1], [38], [40]. Considering eqs 3-5, the 1st term of eq. 2 yields:

$$\frac{\delta I_x}{I_D}|\delta Q_{gr(x)} = \frac{\partial Q_{gr(x)}}{\partial Q_{imp}}\frac{\delta Q_{imp}}{Q_{gr(x)}} = \frac{k|V_c|}{k|V_c|+C}\frac{\delta Q_{imp}}{Q_{gr(x)}} \quad (6)$$

$\mu_{ueff}$ variations are related to $\delta Q_{imp}$ as [28, eqs 3-5]:

$$\frac{\partial \mu_{ueff}}{\partial Q_{imp}} = -\alpha_c \mu_{ueff}^2 \quad (7)$$

allowing to rewrite the 2nd term of eq. 2:

$$\frac{\delta I_x}{I_D}|\delta\beta = \frac{\partial \beta}{\partial Q_{imp}}\frac{\delta Q_{imp}}{\beta} = -\alpha_c \mu_{ueff}\delta Q_{imp} \quad (8)$$

where $\alpha_C$ is the Coulomb scattering coefficient in $V.s/C$ [25], [26] used as a model parameter; $\alpha_C$ has been reported inversely proportional to the square root of inversion charge density ($N_{inv}$) in CMOS [35] but considering it constant from moderate to strong inversion does not significantly err due to its weak dependence on $N_{inv}$ at the specific region [28, Fig. 3]. On the contrary, $\mu_c$ has been experimentally recorded to vary with impurities' density $N_{imp}$ ($\sim 1/N_{imp}$) [32]-[34] but not with $Q_{gr}$ in SL GFETs [30]-[34], in contrast to multilayer ones due to their different density of states and band structure [30]. Thus, $\alpha_c(\sim 1/\mu_c)$ [25], [28], [29]) is also fixed with $Q_{gr}$.

Hence, by using the variance definition of the relative $I_D$ fluctuation $Var[I_D]/I_D^2$ [23, eq. 6], [25, eq. 18] and eqs 1-8, the model reads as (see Appendix (b)):

$$\frac{Var(I_D)}{I_D^2}|\delta Q_{gr(x)},\delta\beta = \frac{e^2 N'_{imp}}{WL^2}\int_0^L \left(\frac{k|V_c|}{(k|V_c|+C)Q_{gr(x)}} \pm \alpha_c\mu_{ueff}\right)^2 dx \quad (9)$$

where the local current fluctuations along the channel are considered uncorrelated. The mobility-related 2nd term in eq. 9 can be either positive or negative depending on whether the impurity is neutral or charged when filled [28]. It is usually positive in Si technologies [28] but negative values have been recorded in some graphene samples due to negative correlation between $Q_{gr}$ and $\mu$ fluctuations [34]. Due to increased inhomogeneities induced by the intense electrolyte-ambient interaction in the specific SG DUT, $N_{imp}$ no longer follows a Poisson distribution [10], leading to a heightened variance. Hence, $N'_{imp}=DN_{imp}$ (in $cm^{-2}$) is defined and used as a model parameter where $D$ is a unitless coefficient equivalent to the $N_{tcoeff}$ (related to traps in [10]) in the range of $10^3\sim 10^6$ (cf. Table I in [10]) and thus: $Var(Q_{imp})=(\delta Q_{imp})^2=e^2N'_{imp}/W\Delta x$ (see Appendix (b)); $D=1$ for solid-gated devices where a Poisson distribution is followed.

Integral variable change from $dx$ to $dV_c$ is given by [40]:

$$\frac{dx}{dV_c} = \frac{-Q_{gr}2L_{eff}}{kg_{vc}}\left(\frac{C_q+C}{C}\right) - \frac{\mu_u}{v_{sat}}\frac{C_q}{C}\frac{|dV_c|}{dV_c} \quad (10)$$

where $g_{vc}$ is a normalized $I_D$ factor, $u_{sat}$ is the saturation velocity, $C_q=k\sqrt{V_c^2+C_1^2}$ is the quantum capacitance with $C_1=U_T.ln(4)$ (in $V$), and thermal voltage $U_T=k_BT/e=25.6\ mV$ at $300\ K$ where $k_B$ is the Boltzmann constant and $T$ the temperature [37, §2.1], [38]. Notice that $kC_1$ product denotes the minimum $C_q$ at CNP ($V_c=0$) due to residual charge effects while away from CNP where $V_c>>U_T$, $C_q\approx k|V_c|$. After considering eq. 3 and eq. 10 in eq. 9, compact eqs 11-17 at the bottom of the next page are derived in terms of $V_c$, where eq. 9 is split into 3 integrals named $I_{DA}$, $I_{DB}$, $I_{DC}$; $I_{DA}$ accounts for $I_D$ variance due to $Q_{gr}$ fluctuations while $I_{DB}$, $I_{DC}$ consider $\mu_{ueff}$ fluctuations additionally. $I_{DA2}$, $I_{DC2}$ mainly denote the VS effect on $I_{DA}$, $I_{DC}$, respectively, through the 2nd term of eq. 10. An additional VS-induced effect on $\mu_{ueff}$ variance shall be considered in a future short-channel analysis where VS cannot be neglected, especially at strong $V_{DS}$. The $R_c$ contribution to $I_D$ variance is:

$$\frac{Var(I_D)}{I_D^2}|\delta R_C = VarR_C\left(\frac{g_m}{2} + g_{ds}\right)^2 \quad (18)$$

where $g_m$ is the extrinsic transconductance, $g_{ds}$ the output conductance [37], while $VarR_c$ is the variance of $R_c$ ($\Omega^2$), used as a model parameter; eq. 18 is adapted by a well-established CMOS model (cf. eq. 2 in [20]).

Thus, the total $I_D$ variance model is derived as:

$$\frac{Var(I_D)}{I_D^2} = \frac{Var(I_D)}{I_D^2}|\delta Q_{gr(x)},\delta\beta + \frac{Var(I_D)}{I_D^2}|\delta R_C \quad (19)$$

For its better experimental calibration, ambipolarity feature is implemented with separate $N'_{impp(n)}$, $\alpha_{Cp(n)}$, $VarRc_{p(n)}$ parameters in p- and n-type regimes, alike $\mu_{p(n)}$, $R_{cp(n)}$ IV model parameters in [38]. $N_{imp}$ differs between p- and n-type devices due to distinct nature of hole (p) and electron (n) impurities [23]. Hence, dissimilar p- and n-type $\mu_c$ (and consequently $\alpha_c$) values have been recorded in SL GFETs [30, Fig. 1a] due to the aforementioned $\mu_c\sim 1/N_{imp}$ relation; $\mu_c$ is the dominant low-field mobility term, mainly defining $\mu_{ueff}$ at low $V_{GS}$ [30]–[32]. The model is valid for both positive and negative $V_{DS}$ [38] while since it is implemented in Verilog-A, it can be easily integrated as an additional module to our GFET compact model [37]-[40].

### III. RESULTS-DISCUSSION

Fig. 2a, 2c, 2e, respectively, present the measured $I_D$ from all the available samples of the *100x100*, *50x50* and *20x20 μm/μm* GFETs, which are afterwards used for the $I_D$ variance model validation. As a first step, basic IV transport parameters ($\mu$, $V_{G0}$, $u_{sat}$, $\Delta$, $\Theta_{int}$, $R_C$) were extracted from the log-mean of the experimental $I_D$ of the three types of GFETs under test (red markers in Fig. 2b, 2d, 2f, respectively) according to [39] (cf. Table 1); $\Delta$ is a residual-charge related parameter [37], [38] and $u_{sat}$ is insignificant for the long-channel DUTs. The *IV* ambipolarity feature of the model introduced in [38], has been applied for better accuracy as asymmetries have been recorded between n- and p-type branches. The consistency of the mean-value (typical-case) *IV* model vs. mean-value measured *IV* data is remarkable for all DUTs at $V_{DS}=50\ mV$. The latter is also confirmed from $g_m$ and channel resistance $R_{ch}=V_{DS}/I_D$ model-measurements accurate fittings (cf. Fig. 2g, 2h, respectively). Next, the standard deviation of $I_D$ has been calculated for both measurements ($\sigma_{meas}[I_D]=\sqrt{\sum(I_{D(i)}-I_{Dmean})/N}$) where $N$ is the number of GFET samples for each geometry, $I_{D(i)}$, $I_{Dmean}$ are the measured $I_D$ for each sample and the log-mean $I_D$ from all measured samples, respectively, at each $V_{GS}$ point) and model ($\sigma_{sim}[I_D]=I_D\sqrt{(Var[I_D]/I_D^2)}$ where the radicand is directly derived from eq. 19 for each $V_{GS}$ point) and plotted (cf. insets in Fig. 2). Models fit precisely the experiments for all DUTs while $\sigma$ presents a deep minimum at CNP. $\pm 3\sigma$ worst cases are also calculated and depicted for all DUTs (cf. Fig. 2b, 2d, 2f) with exceptional agreement between the model and the measurements. Notice the slight difference between mean and $\pm 3\sigma$ values near CNP due to the minimum $\sigma$ there. The latter can be explained in terms of the domination of fixed (and less prone to fluctuations) residual charge over net charge at CNP [37]-[40].

Both experimental and modeled $Var[I_D]/I_D^2$ are displayed in Fig. 3 vs. $V_{GS}$ (left plots) and $I_D$ (right plots) where model lines follow the measured markers consistently for all DUTs. Channel -due to $\delta Q_{imp}$- (eqs 11-17) and $R_c$ (eq. 18) model contributions are also shown separately, revealing that $Var[I_D]/I_D^2$ is dominated by the channel part for all GFETs in most bias regimes. The contribution from $R_c$ variations is strongest for the *100x100 μm/μm* GFET at strong p-type bias (cf. Fig. 3a-3b) since due to highest doping (GFET with the higher $V_{G0}$ as shown in Table I), the specific GFET extends to the strongest p-type operation where contact resistance effects are more intense (cf. Fig. 2); n-type is merely included in the measurements. *20x20*, *50x50 μm/μm* GFETs have similar $V_{G0}$ (and consequently doping), and a modest $R_c$ effect on $I_D$ variance at strong p-type regime is recorded (cf. Fig. 3c-3f), which is slightly more intense for the narrower GFET as $R_c$ is known to be inversely proportional to $W$ (cf.

$$I_{DA} = \frac{e^2 N'_{imp}}{WL^2}\left[\int_{V_{cd}}^{V_{cs}}\left(\frac{1}{Q_{gr}}\right)^2 \left(\frac{k|V_c|}{k|V_c|+C}\right)^2 \frac{Q_{gr}2L_{eff}}{kg_{vc}}\left(\frac{C_q+C}{C}\right)dV_c - \int_{V_{cs}}^{V_{cd}}\left(\frac{1}{Q_{gr}}\right)^2\left(\frac{k|V_c|}{k|V_c|+C}\right)^2 \frac{\mu_u}{v_{sat}}\left(\frac{C_q}{C}\right)|dV_c|\right] = \frac{e^2 N'_{imp}}{WL^2}[I_{DA1}+I_{DA2}] \quad (11)$$

$$I_{DA1} = \frac{4L_{eff}}{Cg_{vc}}\int_{V_{cd}}^{V_{cs}}\frac{V_c^2(C+k\sqrt{V_c^2+C_1^2})}{(V_c^2+\alpha/k)(k|V_c|+C)^2}dV_c = \frac{4L_{eff}}{Cg_{vc}}\left\{\frac{1}{\varphi_1}\left[\frac{\mp C^3}{k}\mp C^2\gamma_1\right]+\frac{1}{\varphi_2}\left[\frac{C^2}{k}tanh^{-1}\left(\frac{V_c}{\gamma_1}\right)\pm\frac{C^3}{k\gamma_2}tanh^{-1}\left(\frac{\gamma_2}{\gamma_1\gamma_3}\right)\right]+\frac{1}{\varphi_3}\left[-\frac{\sqrt{\alpha}C\gamma_4}{\sqrt{k}}tan^{-1}\left(\sqrt{\frac{k}{\alpha}}V_c\right)\mp\right.\right.$$

$$\left.\left.2\alpha C\sqrt{\kappa}\gamma_5\, tan^{-1}\left(\frac{\sqrt{\kappa}\gamma_1}{\gamma_5}\right)+tanh^{-1}\left(\frac{V_c}{\gamma_1}\right)(2\alpha C^2-\alpha\gamma_3)+\sqrt{\alpha}\gamma_4\gamma_5\, tanh^{-1}\left(\frac{V_c\gamma_5}{\sqrt{\alpha}\gamma_1}\right)\pm 2\alpha C\gamma_2\, tanh^{-1}\left(\frac{\gamma_2}{\gamma_1\gamma_3}\right)\mp \alpha C^2\, ln\left(\frac{(C\pm kV_c)^2}{\alpha+kV_c^2}\right)\right]\right\}\bigg|_{V_{cd}}^{V_{cs}} \quad (12)$$

$$I_{DA2} = \frac{4\mu_u}{Cu_{sat}}\int_{V_{cs}}^{V_{cd}}\frac{V_c^2}{(V_c^2+\alpha/k)^2}\frac{k\sqrt{V_c^2+C_1^2}}{(k|V_c|+C)^2}|dV_c| = \frac{4\mu_u}{Cu_{sat}}\left|\left\{\frac{1}{\varphi_3}\left[\frac{\mp C^2 k^2 \gamma_1}{C\pm kV_c}\mp\alpha Ck^2\frac{\gamma_1}{\gamma_6}-\frac{k^2 V_c \gamma_1 \gamma_3}{2\gamma_6}\pm\frac{\alpha C k^{\frac{3}{2}}}{\gamma_5}tan^{-1}\left(\frac{\sqrt{\kappa}\gamma_1}{\gamma_5}\right)\pm C^2 k\, tanh^{-1}\left(\frac{V_c}{\gamma_1}\right)-\right.\right.\right.$$

$$\left.\left.\frac{b^2 k^2 \gamma_4}{2\sqrt{\alpha}\gamma_5}tanh^{-1}\left(\frac{V_c\gamma_5}{\sqrt{\alpha}\gamma_1}\right)\pm\frac{C^3 k}{\gamma_3}tanh^{-1}\left(\frac{\gamma_2}{\gamma_3\gamma_1}\right)\right]+\frac{1}{\varphi_4}\left[\pm 2Ck^{\frac{3}{2}}\gamma_4\gamma_5\, tan^{-1}\left(\frac{\sqrt{\kappa}\gamma_1}{\gamma_5}\right)+C^2 k\gamma_7\, tanh^{-1}\left(\frac{V_c}{\gamma_1}\right)-2C^2 k\gamma_4\, tanh^{-1}\left(\frac{V_c}{\gamma_1}\right)-\right.\right.$$

$$\left.\left.\frac{C^2 k\gamma_5\gamma_7}{\sqrt{\alpha}}tanh^{-1}\left(\frac{V_c\gamma_5}{\sqrt{\alpha}\gamma_1}\right)+2CK\gamma_3\, tanh^{-1}\left(\frac{\gamma_2}{\gamma_3\gamma_1}\right)\right]\right\}\bigg|_{V_{cs}}^{V_{cd}}\right| \quad (13)$$

$$I_{DB} = \frac{e^2 N'_{imp}}{WL^2}\int_0^L (\alpha_c \mu_{ueff})^2 dx = \frac{e^2 N'_{imp}}{WL}(\alpha_c \mu_{ueff})^2 \quad (14)$$

$$I_{DC} = \frac{e^2 N'_{imp}}{WL^2}\left[\int_{V_{cd}}^{V_{cs}}\frac{2k|V_c|\alpha_c \mu_{ueff}}{(k|V_c|+C)Q_{gr}}\frac{Q_{gr}2L_{eff}}{kg_{vc}}\left(\frac{C_q+C}{C}\right)dV_c-\int_{V_{cs}}^{V_{cd}}\frac{2k|V_c|\alpha_c\mu_{ueff}}{(k|V_c|+C)Q_{gr}}\frac{\mu_u}{v_{sat}}\left(\frac{C_q}{C}\right)|dV_c|\right] = \frac{e^2 N'_{imp}}{WL^2}[I_{DC1}+I_{DC2}] \quad (15)$$

$$I_{DC1} = \frac{4\alpha_c\mu_{ueff}L_{eff}}{Cg_{vc}}\int_{V_{cd}}^{V_{cs}}|V_c|\frac{k\sqrt{V_c^2+C_1^2}+C}{(k|V_c|+C)}dV_c = \frac{2\alpha_c\mu_{ueff}L_{eff}}{k^2 Cg_{vc}}[2Ck|V_c|\mp 2C\kappa\gamma_1+k^2 V_c\gamma_1 \mp 2C(C+\gamma_3)\,ln(C\pm kV_c)+(2C^2+C_1^2 k^2)\,ln(V_c+$$

$$\gamma_1)\pm 2C\gamma_3\, ln(\gamma_2+\gamma_3\gamma_1)\big]_{V_{cd}}^{V_{cs}} \quad (16)$$

$$I_{DC2} = \frac{4k\alpha_c\mu_{ueff}\mu_u}{Cu_{sat}}\int_{V_{cs}}^{V_{cd}}\frac{|V_c|\sqrt{V_c^2+C_1^2}}{(k|V_c|+C)(V_c^2+\alpha/k)}|dV_c| = \frac{4k\alpha_c\mu_{ueff}\mu_u}{Cu_{sat}}\left|\left\{\frac{1}{\varphi_2}\left[\mp\frac{C\gamma_5\sqrt{\varphi_2}}{\sqrt{k}}tan^{-1}\left(\frac{\sqrt{\kappa}\gamma_1}{\gamma_5}\right)+\alpha\, tanh^{-1}\left(\frac{V_c}{\gamma_1}\right)+\right.\right.\right.$$

$$\left.\left.\frac{C^2}{k}tanh^{-1}\left(\frac{V_c}{\gamma_1}\right)-\sqrt{\alpha}\gamma_5\, tanh^{-1}\left(\frac{V_c\gamma_5}{\sqrt{\alpha}\gamma_1}\right)\pm C\gamma_3\, tanh^{-1}\left(\frac{\gamma_3}{\gamma_3\gamma_1}\right)\right]\right\}\bigg|_{V_{cs}}^{V_{cd}}\right| \quad (17)$$

with $\varphi_1=\gamma_2(C\pm kV_c)$, $\varphi_2=\alpha k+C^2$, $\varphi_3=\varphi_2^2$, $\varphi_4=\varphi_3^2$, and $\gamma_1=\sqrt{(V_c^2+C_1^2)}$, $\gamma_2=C_1^2 k\mp CV_c$, $\gamma_3=\sqrt{(C^2+C_1^2 k^2)}$, $\gamma_4=\sqrt{(C^2-\alpha\kappa)}$, $\gamma_5=\sqrt{(\alpha-C_1^2 k)}$, $\gamma_6=\alpha+k V_c^2$, $\gamma_7=C^2-3\alpha k$, and $V_{cs(d)}$ is $V_c$ at S(D). ($\varphi 1$, $\varphi 2$, $\varphi 3$, $\varphi 4$, $\gamma 1$, $\gamma 2$, $\gamma 3$, $\gamma 4$, $\gamma 5$, $\gamma 6$, $\gamma 7$ are auxiliary functions used to shorten the equations and improve the readability).

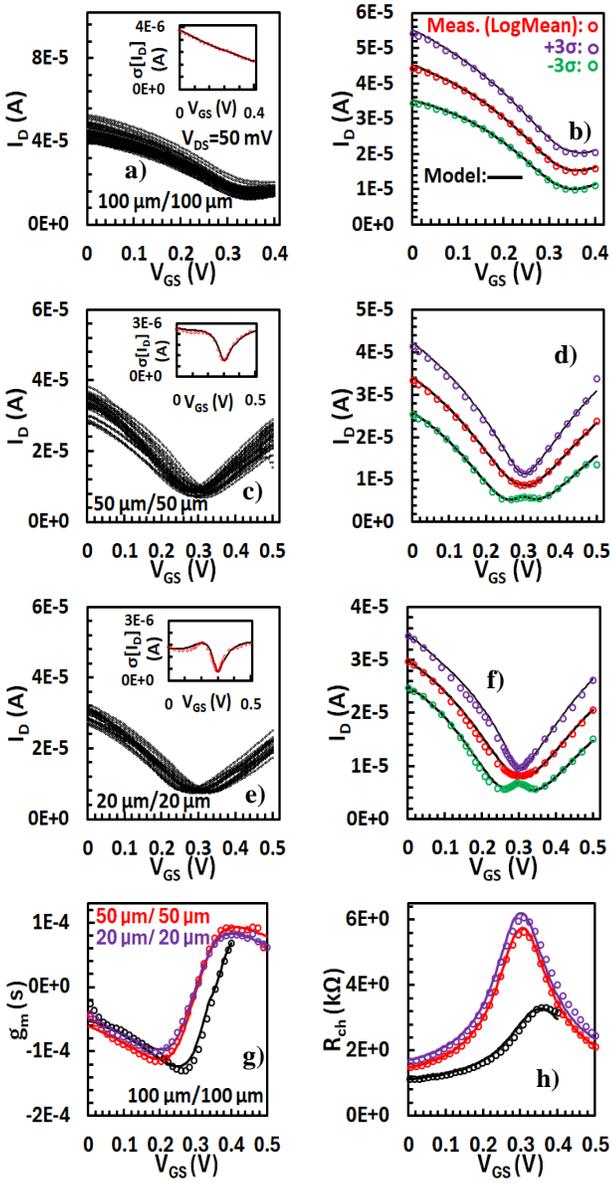

Fig. 2. Transfer characteristics for GFETs with a, b) width $W=100~\mu m$ and length $L=100~\mu m$, c, d) $W=50~\mu m$ and $L=50~\mu m$, and e, f) $W=20~\mu m$ and $L=20~\mu m$. All measured samples are shown in (a), (c), (e) while log-mean and $\pm 3\sigma$ values in (b), (d), (f), respectively. Insets in left column plots display the standard deviation of drain current $\sigma[I_D]$ vs. gate voltage $V_{GS}$. Transconductance $g_m$ and channel resistance $R_{ch}$ vs. $V_{GS}$ for all GFETs are also depicted in (g), (h), respectively. Markers: measurements, lines: model.

TABLE I
$I_D$ AND $I_D$ VARIANCE EXTRACTED PARAMETERS

| Parameter | Units | 100x100 | 50x50 | 20x20 |
|---|---|---|---|---|
| $\mu$ | $cm^2/(V \cdot s)$ | n: 7500 p: 10000 | n: 5216 p: 6300 | n: 5000 p: 5927 |
| $V_{G0}$ | $V$ | 0.331 | 0.278 | 0.283 |
| $u_{sat}$ | $m/s$ | $9 \cdot 10^4$ | $9 \cdot 10^4$ | $9 \cdot 10^4$ |
| $\Delta$ | $meV$ | 42 | 30 | 30 |
| $\Theta_{int}$ | $V^{-1}$ | 3.7 | 2 | 2 |
| $R_c$ | $\Omega$ | n, p: 240 | n, p: 300 | n, p: 385 |
| $N'_{imp} = D \cdot N_{imp}$ | $10^{15} \cdot cm^{-2}$ | n: 12.3 p: 1.47 | n: 210 p: 610 | n: 25 p: 220 |
| $\alpha_C$ | $kV~s/C$ | n: 6 p: 15 | n: 1.1 p: 0.1 | n: 1.05 p: -0.315 |
| $VarR_c$ | $\Omega^2$ | n: 1 p: 6000 | n: 4000 p: 6500 | n: 3000 p: 7000 |

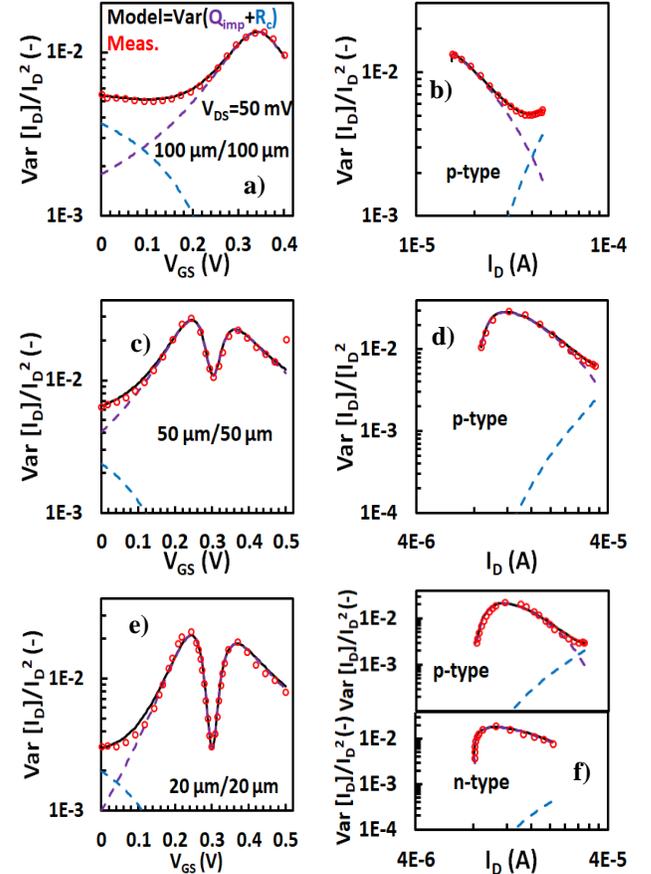

Fig. 3. $Var[I_D]/I_D^2$ vs. $V_{GS}$ (a, c, e) and drain current $I_D$ (b, d, f) for GFETs with a, b) $W=100~\mu m$ and $L=100~\mu m$, c, d) $W=50~\mu m$ and $L=50~\mu m$, and e, f) $W=20~\mu m$ and $L=20~\mu m$. P-type region is shown in (b), (d) while both p- (upper) and n-types (bottom) in (f). Markers: measurements, solid lines: model, dashed lines: Charged impurities $Q_{imp}$ (magenda) and series resistance $R_c$ (blue) contributions.

Table I). A minor $R_c$ effect is also observed at the strong n-type bias regime of the $20x20~\mu m/\mu m$ GFET (cf. Fig. 3f). In general, channel $\delta Q_{imp}$-induced $Var[I_D]/I_D^2$ is exclusively responsible for the recorded M-shape, while $R_c$ can only contribute to high carrier density regions, similarly to $1/f$ noise [26, §6], [37, §5]. The extracted $I_D$ variance model parameters are listed in Table I for every DUT. The extracted $N'_{imp}$ is much higher than the corresponding $1/f$ noise parameter in [10]; similar differences are noticed between [23], [27] for CMOS and [25], [29] for organic FETs, probably because $I_D$ variance is investigated at the DC level as a spatial effect while noise is a temporal effect [23]. Besides, the best measurements-model fitting for the $20x20~\mu m/\mu m$ GFET, is achieved with a negative $\alpha_C$ value at p-type region, confirming the arguments in [34].

For a better comprehension of Coulomb scattering-related $I_D$ variance model ($I_{DB,C}$) and its dissimilarities with Si and organic technologies, simulated channel $Var[I_D]/I_D^2$ is presented in Fig. 4a for the $50x50~\mu m/\mu m$ DUT where identical p- and n-type $\mu$, $N'_{imp}$, $\alpha_C$ parameters were selected ($\mu=\mu_p$, $N'_{imp}=N'_{impp}$ from Table I and $\alpha_C=0, 0.1, 0.5, 1~kV.s/C$) to avoid irregularities. In CMOS [23, Fig. 8, 9] and organic FETs [25, Fig. 4a-c], a peak plateau of channel $Var[I_D]/I_D^2$ is observed at low $I_D$ regime which decreases and reaches its minimum value towards high $I_D$ regime. The 1st term ($\Delta Q/Q$) of eq. 2 is mainly responsible for this plateau as it gets maximum for low $I_D$ [23], [25], similarly to $1/f$ noise case [26, Fig. 6.10], and dominates over the 2nd Coulomb scattering-related term ($\Delta\beta/\beta$) of eq. 2. Thus, $\Delta\beta/\beta$ term contributes only at high $I_D$ regime where $\Delta Q/Q$ drops. On the contrary in GFETs, $Var[I_D]/I_D^2$ presents an M-shape with lowest values at CNP and at strong p- and n-type regions, and maximum ones at the peak points of M-shape. This is also analogous to the $1/f$ noise model, where the M-shape is attributed to $\Delta Q_{gr}/Q_{gr}$ [37, §5]. Consequently, $I_{DB,C}$ has a stronger effect in the regimes where $\Delta Q_{gr}/Q_{gr}$ is less significant, which are the minimum points of $Var[I_D]/I_D^2$ (CNP and strong p(n)-type regions), as presented in Fig. 4a. The latter is also confirmed by the percentage variation of $Var[I_D]/I_D^2$ for $\alpha_C=0.1, 0.5, 1\ kV.s/C$ to $\alpha_C=0\ kV.s/C$, which is also plotted in Fig. 4b. Another important observation is the effect of $\Theta_{int}$ parameter in $I_{DB,C}$ models which is induced through the $\alpha_C.\mu_{ueff}$ product in eq. 9. Note that $\Theta_{int}$ is in the denominator of $\mu_{ueff}$ and its effect is responsible for the degradation of $\mu$ at high density regimes [39] (see Appendix (a)). Hence, de-activating mobility degradation effect by forcing $\Theta_{int}=0\ V^{-1}$ (dashed lines in Fig. 4a) for $\alpha_C=1\ KV.s/C$, results in an increase of $Var[I_D]/I_D^2$ towards strong p-, n-type regions due to increased $\alpha_C.\mu_{ueff}$ product.

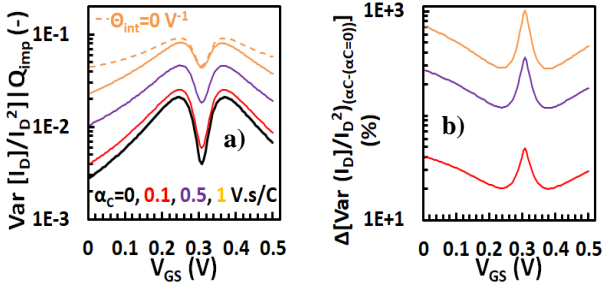

**Fig. 4.** a) Normalized drain current variance $Var[I_D]/I_D^2$ due to $Q_{imp}$ and b) its relative fluctuation from Coulomb scattering coefficient $\alpha_C=0$ to $\alpha_C=0.1$ (red), $0.5$ (magenta) and $1$ (orange) $KV.s/C$, respectively, vs. $V_{GS}$ for GFETs with $W=50\ \mu m$ and $L=50\ \mu m$. Solid lines: model, dashed lines: mobility degradation coefficient $\Theta_{int}=0\ V^{-1}$.

## IV. CONCLUSIONS

A physics-based GFET $I_D$ variance model is proposed for the first time here, which accurately describes the bias-dependence of $I_D$ variability after validation with experiments from adequate statistical samples of SG GFETs at a wide range of bias conditions. The followed methodology reveals an equivalence between $I_D$ variance and $1/f$ noise models, since identical physical channel mechanisms are identified to originate them, namely the fluctuations of both $Q_{gr}$ and Coulomb scattering-related $\mu_{ueff}$ induced by $Q_{imp}$ deviations. The validity of our approach is based on the consideration of local current fluctuations which result in solid derivations even for non-uniform channel conditions. A $R_c$ variance feature is also included, which was found to contribute at very high charge density regimes. $Q_{imp}$ dominates in the rest of the transport regime with $\mu_{ueff}$ more critical at CNP and at strong p(n)-type regions.

The proposed straightforward model ensures fast simulations with a small computational burden in contrast to more complicated solutions such as Monte-Carlo simulations. Such a variance model can be beneficial for industry and academic device technology groups to predict from worst to best performance scenarios of a device, as well as for circuit designers, as a circuit can be tested within a practical $I_D$ range, related to the actual device, and hence, tolerance design rules can be foreseen.

## APPENDIX

a) By considering i) $\Delta R=\Delta x/(W|Q_{gr}|\mu_{diff})$ [40, eq. 11], ii) $G_{CH}=W|Q_{gr}|\mu_{ueff}/L_{eff}$ [40, eq. 10, 14] and iii) eq. 2 in [38] (which proves that $\mu_u/\mu_{ueff}=L_{eff}/L$), eq. 1 is derived; $\mu_{ueff}=\mu_u/(1+|E_x|/E_C)$ with $E_x$ and $E_C$ the horizontal and critical electric fields, respectively [40], $\mu_u=\mu/(1+\Theta_{int}\sqrt{(V_o^2+(V_{GS}-V_{G0})^2)})$ which includes mobility degradation due to vertical field with $V_o$ a residual charge related voltage [39], and $\mu_{diff}=\mu_{ueff}+\partial\mu_{ueff}/\partial E_x.E_x$ [26, §9.4.1], [40].

b) Eq. 9 can be derived as in [23], [25] by using eqs 1-8:

$$\frac{Var(I_D)}{I_D^2}\Big|_{\delta Q_{gr(x)},\delta\beta} = \sum \overline{\left(\frac{\delta I_d}{I_D}\right)^2} = \lim_{\Delta x \to 0}\sum \overline{\left(\frac{\Delta x}{L}\frac{\delta I_x}{I_D}\right)^2} =$$
$$\frac{1}{L^2}\int_0^L \Delta x \left(\frac{\delta I_x}{I_D}\right)^2 dx = \frac{1}{L^2}\int_0^L \Delta x \left(\frac{k|V_c|}{(k|V_c|+C)Q_{gr(x)}} \pm \alpha_c\mu_{ueff}\right)^2 (\delta Q_{imp})^2 dx \quad (A1)$$

$Q_{imp}$ variance is $Var(Q_{imp})=(\delta Q_{imp})^2=e^2N_{imp}/W\Delta x$ [23, eqs 9-11] (identical to the $1/f$ noise trapped charge density [26, eq. 6.65]) after assuming a Poisson distribution where the variance of the total number of impurities equals to its mean value $n_{imp}=W.\Delta_x.N_{imp}$ with $N_{imp}$ the impurities density ($cm^{-2}$).